\documentstyle[11pt,aas2pp4]{article}
\begin{document}
\def\etal{\it et al. \rm }
\title{Color Evolution from z=0 to z=1}

\author{Karl D. Rakos\altaffilmark{1}}
\affil{Inst. for Astronomy, Univ. of Vienna}
\altaffiltext{1}{Visiting Astronomer, Kitt
Peak National Observatory, National Optical Astronomy Observatories, which
is operated by the Association of Universities for Research in Astronomy,
Inc. (AURA) under cooperative agreement with the National Science
Foundation.}

\author{James M. Schombert\altaffilmark{1,2}}
\affil{Infrared Processing and Analysis Center\\Jet Propulsion
Laboratory\\California Institute of Technology}
\altaffiltext{2}{Present Address: Astrophysics Divison, Code SZ, NASA
Headquarters, Washington, D.C. 20546}

\begin{abstract}
Rest frame Str\"omgren photometry (3500{\AA}, 4100{\AA}, 4750{\AA} and
5500{\AA}) is presented for 509 galaxies in 17 rich clusters between $z=0$
and $z=1$ as a test of color evolution. Our observations confirm a strong,
rest frame, Butcher-Oemler effect where the fraction of blue galaxies
increases from 20\% at $z=0.4$ to 80\% at $z=0.9$.  We also find that a
majority of these blue cluster galaxies are composed of normal disk or
post-starbursts systems based on color criteria. When comparing our colors
to the morphological results from HST imaging, we propose that the blue
cluster galaxies are a population of late-type, LSB objects who fade and
are then destroyed by the cluster tidal field. After isolating the red
objects from Butcher-Oemler objects, we have compared the mean color of
these old, non-star forming objects with SED models in the literature as a
test for passive galaxy evolution in ellipticals. We find good agreement
with single burst models which predict a mean epoch of galaxy formation at
$z=5$. Tracing the red envelope for ellipticals places the earliest epoch
of galaxy formation at $z=10$.
\end{abstract}


\section{INTRODUCTION}

Observational astronomy has one major advantage over any other physical
science, the phenomenon of lookback time. The combination of enormous
distances plus the finite speed of light permits the study of the behavior
of objects in our distant past and, with respect to extragalactic
astronomy, this allows direct observation into the evolutionary history of
galaxies, as opposed to just deducing the past based on their contents at
the current epoch. This furnishes data for the investigation of the
evolution of stellar populations, star formation history and the
conditions of galaxy formation, all relatively new fields as our telescope
collecting power has increased in recent years.  Unfortunately, the study
of distant galaxies is inhibited by several difficulties primary being
that their great distances imply small sizes and faint apparent
luminosities. In addition, large distance also means the redshifting of
regions of interest (e.g. the near-blue) into the technically and
observationally troublesome near-IR.

The study of color evolution of galaxies has divided into three parts in
recent years.  Foremost are optical and near-IR studies of extremely high
redshift galaxies, mostly radio selected systems, to probe the conditions
immediately after the time of galaxy formation (Eisenhardt and Chokshi
1990, McCarthy, Perrson and West 1992). The main purpose of these studies
has been to search for protogalaxies, but they also place tight
constraints on the the star formation history of galaxies. Second are
color selected galaxy surveys in broadband colors or the 4000\AA\ break
which have been used to test spectrophotometric model predictions out to
redshifts of two (Hamilton 1985, Eisenhardt and Lebofsky 1987). Lastly are
the numerous photometric studies on the fraction of blue galaxies in
clusters (Butcher and Oemler 1984, Dressler, Gunn and Schneider 1985)
relating to the rapid changes in cluster populations at only modest
redshifts (0.2 to 0.5).

This study is a photometric analysis of galaxy cluster populations out to
a redshift of one (10 Gyrs ago, $H_o = 50$ km sec$^{-1}$ Mpc$^{-1}$,
$q_o=0$). We attempt a compromise between a full spectral analysis of
distant clusters, versus simple, K-corrected broadband colors, by using
narrow band blue filters ``redshifted" to the cluster redshift
($\lambda_{obs}=\lambda_o(1+z)$). This use
of narrow bandpasses, rather than direct spectroscopy, produces some loss
in spectral resolution but gains in increased S/N per object and coverage
of the entire cluster per exposure.  We present new photometry for 8
clusters from $z=0.6$ to $z=1$ (the limit of ground based, optical
photometry). When combined with our previous results from our zero, low
($z=0.2$) and intermediate ($z=0.4$) redshift samples (Fiala, Rakos and
Stockton 1986, Rakos, Fiala and Schombert 1988, Rakos, Schombert and
Kreidl 1991, Schombert \etal 1993), we can use the results to address four
galaxy evolution questions:  1) the change in the fraction of red to blue
galaxies with redshift (Butcher-Oemler effect), 2) the nature of blue
cluster galaxies, 3) the mean colors of ellipticals as a function of
redshift (color evolution) and 4) the redshift of galaxy formation.

\section{OBSERVATIONS}

These observations were taken with the KPNO 4m PFCCD using Str\"omgren
$uvby$ filters centered at 3500{\AA}, 4100{\AA}, 4750{\AA} and 5500{\AA}
(hereafter referred to as $uz$, $vz$, $bz$ and $yz$) in the rest frame of
the clusters studied. The sample of high redshift clusters was taken from
the lists of Gunn, Hoessel and Oke (1986) which provided redshifts,
coordinates and finding charts. Each of these filters are approximately
200\AA\ wide (see Figure 1) and were specially designed so that they are
``redshifted" to the cluster wavelength in order to maintain a rest frame
color system.  The KPNO data was taken over three runs in Oct 1990, Jun
1991 and May 1992.  The PFCCD setup with the TE1K chip, a 1024 by 1024 CCD
with a 70\% QE at 8000\AA\ and 0.48 arcsecs per pixel was used on all
runs. A total of over 15 filters ranging from a wavelength of 5500 to
9900{\AA} were used. Flattening was performed by dithering the exposures
of 600 secs and making sky flats.  Most of the data was taken during
bright time, but OH emission dominated the sky counts rather than
scattered moonlight.  Calibration used spectrophotometric standards from
Massey \etal (1988) combined with tabulated filter transmission curves.
Colors and magnitudes were measured using FOCAS and are based on metric
apertures set at 32 kpc for cosmological parameters of $H_o = 50$ km
sec$^{-1}$ Mpc$^{-1}$ and $q_o=0$. Our typical errors were 0.05 in $vz-yz$
and 0.06 in $bz-yz$ for AB(5500)=18 galaxies at the bright end of the high
redshift sample and 0.13 and 0.18 for the faint end of AB(5500)=21. A
greyscale image of our most distant cluster, CL1622.5+2352, at a redshift
of 0.927 is shown in Figure 2.

Since our filters are specific to the redshift of the cluster, we can use
the spectral shape of galaxies around the 4000\AA\ break to
discriminate between stars, foreground and background objects.  Stars and
foreground galaxies will be sampled by the redshifted filters along the
red portions of their spectra producing artificially blue $uz-yz$ and
$vz-yz$ colors. Background galaxies will have their $vz$ fluxes in the
near-UV producing flat spectrum $uz-vz$ colors.  Discriminating between
foreground and background objects is performed using our $mz$ index
described in Paper III. But, we note that our observations are pointed at
distant rich clusters and any foreground contamination will be minimal due
to the small angular size of the cluster and any background contamination
is restricted by the S/N depth of the frames.  In addition, one can see
from inspection of Figure 1 that our filter system avoids all the major
emission lines, such as [OII] at 3727\AA\ or H$\beta$ at 4861{\AA},
associated with either star formation or active galactic nuclei (AGN)
phenomenon, thus preventing the exclusion of unusual galaxy types due to
emission lines. 

Histograms of $bz-yz$ for all the clusters are found in Figure 3. Data for
clusters with redshifts less than 0.5 were taken from Papers I through III
of our series and displayed herein for completeness.  Table 1 presents the
reduced cluster data, again for our new data and all clusters from Papers
I through III.  The table formation is as follows: column 1 is the cluster
id, column2 is the redshift of the cluster taken from the literature,
column 3 is the number of galaxies in the cluster with available colors in
at least three passbands, column 4 and 5 are raw cluster mean colors,
column 6 is the $bz-yz$ color selection for star forming objects (see \S
3.3), column 7 and 8 are the mean cluster colors with the color cut applied
and columns 9 and 10 are the fraction of blue galaxies in each cluster
defined by the indicated color. Cluster errors are 1$\sigma$ errors on the
means for each color. Rather than publishing extensive tables of our
photometry, we have arranged to make the data available by contacting
either author on email (js@ipac.caltech.edu).

\section{DISCUSSION}

The extragalactic meaning of the Str\"omgren colors are described in
detail in our previous papers.  Briefly, the Str\"omgren colors are crude
estimators of recent star formation, mean age and metallicity. The $uz$
filter is centered in the near-UV and measures the amount of recent
high-mass star formation, $vz$ is centered of the CN-Fe blend at
4170\AA\ and is sensitive to metallicity effects and $bz$ plus $yz$ are
centered on regions with no strong spectral features and serve as
continuum measures. In previous papers, we have used $uz-vz$ as a measure
of the 4000{\AA} break, $vz-yz$ as a measure of global metallicity and
$bz-yz$ as a measure of mean age of the underlying stellar population.
However, precise understanding of these colors requires comparison to SED
models with various assumptions (e.g. redshift of formation, mean
metallicity and IMF) since varying star formation histories inhibit a
unique interpretation of the colors. For example, $vz-yz$ colors are only
sensitive to metallicity only for a homogenous age population. Recent star
formation can also strongly influence $vz-yz$ (see Figure 1). To this end,
we have convolved our filters with models from Guiderdoni and
Rocca-Volmerange (1987) and restrict our interpretation to global
properties of the sample and state the model dependence of our results
specifically in our discussion and conclusions.  In addition, we have
focused our analysis on four phenomenon, the changing fraction of blue to
red galaxies in clusters (Butcher-Oemler effect), the nature of this blue
cluster population, the color evolution of red galaxies (assumed to be
the progenitor of present-day ellipticals) and estimating the epoch of
galaxy formation from the red envelope (O'Connell 1987).

\subsection{BUTCHER-OEMLER EFFECT}

The Butcher-Oemler effect is the strongest evidence of direct evolution of
the stellar populations in galaxies that has been discovered to date. In
its simplest form, the Butcher-Oemler effect is the observed increasing
ratio of blue galaxies in a cluster as a function of redshift.
There has long been an expectation that galaxy colors change with redshift
since the gas depletion rates for many galaxy types are less than a Hubble
time. Cessation of star formation naturally leads to a reddening of the
integrated colors of the galaxy as massive blue stars in the underlying
stellar population evolve to the red giant branch. Also, in practical
terms, star formation rates are zero today for ellipticals and S0's,
whereas they must have been non-zero at some point in the past to produce
the current population. Thus, at some past epoch, their mean integrated
colors must have been bluer. However, the Butcher-Oemler result is
surprising because of the extremely rapid change in the fraction of blue
galaxies from nearly zero at the current epoch to approximately 20\% at
only modest redshifts of 0.4, approximately 4 Gyrs ago (Butcher and Oemler
1984).  In contrast, even evolutionary models with late galaxy formation
epochs predict that color changes were minor below a redshift of 0.5.

Our color data is unique from previous cluster studies in that we can
determine the fraction of blue to red galaxies in rest frame colors
without any K-corrections or model dependent parameters. Our color indices
also remove any background or foreground contamination, a major inhibitor
in broadband photometry surveys which require substantial background
corrections or spectroscopic redshift confirmation. Our narrow band colors
are also more finely tuned to test for star formation versus metallicity
effects.  In addition, matching to the cluster redshift eliminates
contamination from emission lines.

The original broadband definition of the fraction of blue to red galaxies,
$f_B$, given by Butcher and Oemler (1984) is the fraction of galaxies 0.2
mags bluer than the mean color of the E/S0 sequence after K-corrections to
the total number of galaxies in the cluster.  As discussed in Paper III, a
sample of nearby spirals and irregulars is used to determine a value of
$bz-yz$ and $vz-yz$ in our filter system that separates star-forming from
quiescent galaxies. From this previous analysis, we have defined the
fraction of blue galaxies, $f_B$, as the ratio of the number of galaxies
bluer than $bz-yz$=0.20 or $vz- yz$=0.40 to the total number of galaxies.
Since the mean $bz-yz$ color of a present-day elliptical is 0.37
(Schombert \etal 1993) and $bz-yz$ maps into $B-V$ in a linear fashion
with a slope of 1.33 (Matsushima 1969), then a cut at $bz-yz$=0.20 is
effectively the same as Butcher and Oemler's 0.2 mag selection. To match
limiting absolute magnitudes from cluster to cluster, we have only used
galaxies with $yz$ mags greater than the magnitude of the 3rd ranked
galaxy plus three (similar to Abell's definition of cluster richness).
Since the lower redshift clusters were imaged on 1 to 2 meter class
telescopes, and the clusters above $z=0.6$ on the KPNO 4m, we found the
completeness of the data in terms of absolute luminosity was effectively
balanced by S/N.  Both $bz-yz$ and $vz-yz$ are used to calculate $f_B$ and
are listed in Table 1.  Our cluster data is shown in Figure 4 along with
the original Butcher and Oemler (1984) data and a dotted line as our
interpretation of the trend discussed below. Errors are $\sqrt{N}$ the
number of members in each cluster.  The value of $f_B$ from either $bz-yz$
or $vz-yz$ color produces basically the same distribution with redshift,
as seen in Figure 4, even though $vz-yz$ is sensitive to metallicity
effects and $bz-yz$ specifically tests continuum colors.  This would
support our claim that foreground galaxies have not significantly
contaminated our sample since foreground galaxies would produce bluer
$vz-yz$ colors relative to $bz-yz$ and background galaxies produce very
red $vz-yz$ colors and flat spectrum $uz-vz$ colors.  In either scenario,
our color selection index, $mz$, would remove these objects from the
sample as demonstrated in Papers I through III.

Our measurements of $f_B$ can be tested by comparison to previous
studies.  For example, our sample has two clusters in common with the
Butcher and Oemler (1984), A370 and CL0024.5+1653, with their values of
$f_B$ of 0.21 and 0.16 respectively. Our values are 0.18$\pm$0.10 and
0.17$\pm$0.10, in good agreement. In addition, one cluster (CL0939.7+4713)
in our sample was observed by HST and published in a morphological/color
study of high redshift clusters (Dressler \etal 1994).  Our color criteria
of $bz-yz<0.20$ corresponds to $g-r<0.02$ in rest frame colors.  Applying a
K-correction of 1.19 for a redshift of 0.04 (Schneider, Gunn and Hoessel
1983) gives a $g-r<1.21$ cutoff for blue galaxies. For galaxies with
morphological classification brighter than $g$=22.0 (100 galaxies), the
Dressler \etal data produces a value of $f_B$ of 0.34, which is well
within the errors of our value of 0.28$\pm$0.10. The morphological
information of the blue cluster galaxies is also insightful.  Of the 100
galaxies, 42 are described as early-type (E,L,S in their system) and 58
are classed as spiral or merger (A,B,C,D,M). Of the blue galaxies
($g-r<1.21$), only 3 are early-type and 33 are late-type. Of the red
galaxies, 39 are early-type and 25 are late-type.  Thus, it appears that
the blue cluster population contains few early-type galaxies (as
expected), but the red population contains a mixture of ellipticals, S0's
and spirals (see discussion below).  

Beyond a redshift of 0.2, the data in Figure 4 shows that $f_B$ increases
steadily to a value of 0.80 by a redshift of 0.9. This change in cluster
populations is dramatic, not only in terms of the rapid pace of galaxy
evolution as first discovered by Butcher and Oemler, but also in the
extent of the blue galaxy population dominates the entire cluster
population at high redshifts.  The largest value of $f_B$ from either
Butcher and Oemler (1984) or Dressler, Gunn and Schneider (1985) was 0.36
at redshifts of 0.4.  The dashed line in Figure 4 represents Butcher and
Oemler's interpretation of the trend with redshift from their sample of
$z<0.4$ clusters.  A second line is draw displaying our interpretation
from the high redshift sample. Our data suggests a slightly more rapid
Butcher-Oemler effect than the lower redshift studies. In fact, many of
our low redshift clusters have higher $f_B$ values than Butcher and
Oemler's clusters at the same redshifts.  The present-day mixture of
spirals in rich clusters is 20\% Sa's, 16\% Sb's and 4\% Sc+Irr's
(Whitmore, Gilmore and Jones 1993). The mean $B-V$ color of these types
are 0.94, 0.87 and 0.69 respectively (Oemler 1991).  The 0.20 mag cutoff
for the calculation of $f_B$ signifies that only galaxies with colors bluer
than $B-V$=0.79 are selected. Thus, only Sc's and Irr's are counted in
present-day clusters which agrees with Butcher and Oemler's value of 0.04
for low redshift clusters. However, our narrow band filters are less
sensitive to reddening effects ($E(bz-yz)=0.22$ versus $E(B-V)=0.32$) and
a 0.10 mag difference in internal reddening is sufficient to include Sa
and Sb type galaxies into our measurements of $f_B$. This would explain
why are values of $f_B$ for $z=0.2$ clusters are, on average, higher than
the mean value from Butcher and Oemler clusters at the same redshift.

The relationship outlined in Figure 4 is not smooth nor clearly linear.
Two clusters (A227 and A2317) display high fractions of blue galaxies at
only $z=0.2$ ($f_B=$ 0.51 and 0.65). One cluster at $z=$0.66
(CL0128.8+0628) has a low value for its redshift of $f_B$=0.25.  The
Butcher-Oemler effect in previous studies has shown similar behavior in
that the observed pattern is more a deficiency of clusters dominated by
red galaxies at high redshift, rather than an overabundance of blue
dominated clusters or a correlated trend with redshift. In other words, at
low redshift one can find clusters with both high and low values of $f_B$,
but at higher redshifts we only find clusters with high $f_B$ values.  One
possible explanation for the scatter at low redshifts is that various
cluster types (irregular to compact) are being selected, whereas at higher
redshifts only the richest, densest clusters are detected and cataloged.
For example, Allington-Smith \etal (1991) finds that the value of $f_B$
varies with cluster luminosity from a value of 0.30 for high luminosity
clusters to 0.05 for low luminosity clusters. This would introduce a bias
in that clusters detected at high redshift are found to be of much higher richness
(i.e. cluster mass) than surveys of nearby clusters (the Scott effect).
This might select clusters with stronger gas gradients at high $z$
relating to higher gas stripping rates or cluster galaxies with later
infall times relating to a longer lasting blue population. This could be
interpreted that the mechanism behind the Butcher-Oemler effect works best
in dense clusters, or the Butcher-Oemler effect takes time to evolve and
the richest clusters are the oldest (Mamon 1986).

\subsection{THE NATURE OF BLUE CLUSTER GALAXIES}

The origin of the blue galaxies that result in the Butcher-Oemler effect
has had two possible interpretations in the literature. The original
papers by Butcher and Oemler propose that the blue cluster galaxies are
unusual due to ongoing star formation. However, a different interpretation
was proposed by Dressler and Gunn (1983) based on spectroscopic
observations of Butcher-Oemler clusters. They found that some of the
Butcher-Oemler galaxies did indeed have strong emission lines indicative
of a protracted period of star formation, but there also exists an unusual
number of objects with post-starburst signatures, such as strong Balmer
absorption features (E+A galaxies), and with AGN features. For example,
the 3C 295 cluster displayed a tenfold increase in the number of galaxies
with high excitation emission lines compared to present-day clusters
(Dressler and Gunn 1983).  Thus, the mystery lies both in the rapid
timescale of the blue population's evolution, as related to the
spectroscopic evidence of starbursts, and the identification of the
ancestors of the blue population in present-day clusters.

The blue cluster population is selected based on $bz-yz$ continuum colors,
although the same population is identified from $vz-yz$ colors. However,
the 4000\AA\ break colors, $uz-vz$, are more relevant to testing the style
and existence of star formation. Figure 5 displays the histograms of
$uz-vz$ and $uz-yz$ color for the sample of clusters with redshifts less
than 0.6 and greater than 0.6. There is no significant change in the
$uz-vz$ colors between the high and low redshifts unlike the sharp change
in mean $vz-yz$ or $bz-yz$ colors from Figure 3. The bottom panels of
Figure 5 display the high redshift sample divided into red and blue
galaxies based on the $bz-yz<0.2$ criteria. Although the blue galaxies
have a slightly bluer median $uz-vz$ color than the red galaxies, the
difference is not what would be predicted from changes in $bz-yz$ or
$vz-yz$ (Schombert \etal 1993).  The same galaxies are easily
distinguished in $uz-yz$. In other words, the blue cluster galaxies
at high redshift have redder $uz-vz$ colors than would be expected from
their continuum colors and higher $uz$ fluxes compared to the red
galaxies.  Redder $uz-vz$ color signals a strong 4000\AA\ break which, in
turn, is also an indication of a hot star component due to recent star
formation. This can be seen by considering a typical old population
spectrum in the 4000 to 5000\AA\ region shown in Figure 1. Any hot, young
star component will increase the near-UV flux as seen in the $uz-yz$
histograms, but will also sharply increase the flux on the red side of the
4000\AA\ break (i.e.  compare the 1.5 Gyr model to the 17 Gyr model in
Figure 1), resulting in redder $uz-vz$ colors compared to a normal
elliptical SED.  We rule out emission lines or AGN activity as a source of
blue colors since our filters are specifically chosen to avoid all the
significant emission features in the near-blue (O[II], H$\beta$, O[III],
etc.). If the blue galaxy population was due primarily to emission
features, then they would not be distinguished in our narrow band colors
as unusual.  Therefore, we interpret the trend in $uz-vz$ colors with
redshift as evidence that the blue galaxies primarily have a young to
intermediate age, post-starburst component dominating their colors rather
than a strong, ongoing star formation episode.

The blue cluster population is a sample of galaxies with recent star
formation, so there are three global possibilities for their evolution
since the same population is absent in present-day clusters.  First,
they have either evolved into some known red cluster galaxy type, such as
cluster ellipticals and S0's. Second, they have faded from view and are
simply not cataloged in our surveys of cluster populations (i.e. are of
low surface brightness galaxies, LSB). Or, third, they are destroyed as
identifiable units.

Considering the last option first, it is difficult to imagine a scenario
where over 80\% of a cluster population ($f_B$ values at $z=0.9$), or
about 1/2 the total cluster luminosity, is destroyed with no remaining
evidence.  There are strong tidal forces to play in rich clusters;
however, interactions with other cluster members or a central cD galaxy
occur at high velocities which is a condition that is not accommodating to
tidal disruption (short interaction times) as much as tidal stripping
(Merritt 1985, Malumuth and Richstone 1984). Even assuming some mechanism
that only disrupts blue galaxies while ignoring red ones (perhaps the blue
galaxies are low density structures that formed late), then the newly
freed stellar population are still bound to the cluster potential. These
stars would produce a luminous halo centered on the cluster core and,
although giant halos have been detected around cD galaxies (Oemler 1976,
Schombert 1988) and they have total luminosities which rival the brightest
cluster members luminosity and not the total cluster luminosity necessary
to explain the missing blue cluster population.

The second option is that blue cluster galaxies have faded from view. As
the upper main sequence of a stellar population is depopulated after a
burst of star formation, the red giant branch grows and the integrated
colors redden. In addition, as high mass stars evolve into low luminosity
white dwarfs, the luminosity per square parsec declines and the mean
surface brightness of the galaxy decreases. For a disk population with a
standard IMF, there is a change in one blue mag arcsec$^{-2}$ for every
0.23 mags change in $B-V$ color and, over a period of 5 Gyrs, a galaxy can
drop 4 to 5 mags from its peak blue luminosity (Arimoto and Yoshii 1987).
The peak luminosity of the blue cluster galaxies must be of order
$M_B=-$19 to $-$20 for the blue population to be detected in the cluster
surveys thus far. This amount of luminosity is only obtained from a
starburst involving $4\times10^9 M_{\sun}$ of material.  However, this
is a strong burst of star formation where a significant fraction of the
mass of the galaxy is turned into stars and, after such an episode, the
total number of stars has not decreased so its final surface brightness
would remain high.  To see this consider a normal sized disk galaxy with
an exponential scale length, $\alpha$, of 2 kpc with total blue mags of
$-$20. Its initial central surface brightness ($\mu_o \propto M_B + 5\, {\rm
log}\, \alpha$) is 20.0 $B$ mag arcsec$^{-2}$ which would fade to 24.0
$B$ mag arcsec$^{-1}$ in 5 Gyrs, still quite visible in present-day
cluster surveys.  In order for fading to be plausible, consider a larger
galaxy with an $\alpha=10$ kpc and the same absolute magnitude. It would
have an initial central surface brightness of 22.5 $B$ mag arcsec$^{-2}$
which would fade to 26.5 $B$ mag arcsec$^{-2}$.  If we assume there were
no high surface brightness bulge components (e.g., galaxy type Sc or
later), then such a galaxy would be invisible on the Palomar Sky Survey
prints and missing from any galaxy catalog.

This proposed LSB population would also explain the high fraction of AGN
activity in high redshift clusters (Dressler and Gunn 1983). In a study of
nearby, giant LSB galaxies, Knezek and Schombert (1994) found that 60\%
have low luminosity AGN activity. The weak emission is assumed to be due
to the low surface density of gas in the cores of these systems, making a
deficiency in fuel for the central engine. If the same event which
triggers star formation in the blue cluster population also increases the
core gas density, then the hidden AGN would increase in luminosity and the
higher AGN fraction observed by Dressler and Gunn would be realized.

Unfortunately, there is presently no observational support for a hidden
LSB population in present-day clusters. Photographically enhanced surveys
of the Virgo cluster (Binggeli, Sandage and Tammann 1985, Impey, Bothun
and Malin 1988) have achieved a depth of $\mu_{lim} \approx 27$ $B$ mag
arcsec$^{-2}$ and have not detected a population of large, LSB disk
galaxies. Field surveys for LSB galaxies (Schombert \etal 1992) have found
numerous examples of LSB counterparts to normal disk galaxies (scale
length $\alpha$ of 2 to 4 kpc) and large ($\alpha > 10$ kpc) Malin
galaxies, but none in a cluster environment (see Bothun \etal 1993).
Analysis of the structure of LSB galaxies demonstrates that their low
luminosity densities reflect low surface mass densities and, thus, their
is an expectation that such systems would not survive in a cluster
environment (McGaugh 1992).

The remaining possibility for the fate of the blue cluster population is
that they have evolved into some other kind of galaxy type. This galaxy
type would also have to be red so as to reconcile the current distribution
of galaxy colors in clusters to those in the past. Although there are
highly reddened, dust-rich Sa's in clusters, the dominate red galaxy types
are ellipticals and S0's. Ellipticals are poor candidates for the blue
cluster population since they are composed of a single burst population of
at least 12 Gyrs old with no evidence of recent star formation (Wyse
1985) plus have the correct morphological fraction from low to high
redshift.  In the previous sections, we have shown that a large fraction
of the cluster members have star forming colors by $z=0.9$.  Since 40\% of
present-day cluster galaxies are S0's and 40\% are spirals, the change
from 20\% at $z=0.2$ to 80\% at $z=0.9$ is suggestive that the increasing
number of blue galaxies in clusters are star-forming S0's and early-type
spirals. The decline in blue galaxies to the present epoch would then
represent the gradual halt in star formation due to gas depletion.  The
later passive evolution of the stellar population from a young, blue one
to an old, red one would result in a quiescent S0.  The rapid reddening of
blue cluster galaxies is also not unexpected in the context of
spectrophotometric models. For example, Ellis (1988) showed that a 10\%
burst (i.e. a strong burst) on top of an old population can evolve in less
than one Gyr to a galaxies whose colors are indistinguishable from a 16
Gyr population and blue galaxies at $z=0.5$ have sufficient time to evolve
into the red galaxies. A plausible scenario is one where large bulge S0's
have a burst or past episode of star formation in their disks. For
redshifts from 0.4 to 0.9, the brightness of the disk dominates and the
integrated color of the galaxy is blue. As the disk fades, the bulges
dominates and the galaxies quickly becomes red. This hypothesis is
supported by the near-IR colors of S0 disks which indicate that they have
had their last episode of star formation 4 Gyrs ago or redshifts of 0.3
(Bothun and Gregg 1990).

It is tempting, then, to envision an evolutionary progression where
spirals are converted to S0's and various mechanisms have been proposed
over the past decades using ram pressure stripping or other gas depletion
methods (Spitzer and Baade 1951, Gunn and Gott 1972, Larson, Tinsley and
Caldwell 1980). However, the single greatest barrier to relating S0's to
spirals has been that the distribution of bulge to disk ratios (B/D) for
S0's is extremely difference from that of spirals (Dressler 1980). The
Hubble sequence is also a sequence of increasing B/D from Sc to S0, where
B/D is defined either by isophotal size or relative luminosity. If the
Butcher-Oemler effect is a progression of spirals exhausting their gas
supply with star formation to become red S0's, then the progenitors must
be large B/D, early-type spirals.  Whitmore, Gilmore and Jones (1993)
propose that the B/D ratio is also an indicator of formation time, with
larger B/D galaxies forming first. Then, the succession from S0 to Sc is a
chain of formation epochs where the older galaxies (i.e. future S0's) run
out of gas first.  Thus, the cluster Sa's of today are the S0's of
tomorrow and the separation of B/D between Hubble types is maintained, not
by converting late-type spirals into S0's, but by merely adding
early-type, large B/D galaxies to the S0 class in a steady fashion. This
slowly shifts the distribution of B/D for S0's as a function of redshift,
but maintains their high mean B/D value as compared to star-forming,
gas-rich spirals in the cluster. If this conversion process is responsible
for the Butcher-Oemler effect, then the blue galaxies in clusters at high
redshift should be large B/D spirals where their blue component is a
bright, star-forming disk.

Further insight to this scenario can be found if we return to the HST
results on CL0939.7+4713 at $z=0.40$ (Dressler \etal 1994) and examine the
morphological types of the blue galaxies. Of the 100 galaxies brighter
than $r=22$ with morphological classification, 20 are ellipticals, 22 are
S0's (type S or L) and 58 are late-type galaxies (Sa's through Sd's plus
seven interacting/merger systems). In contemporary clusters, the mean
morophological mixture is 20\% E's, 40\% S0's and 40\% spirals and
irregulars (Oemler 1991) so already we can see that there is a deficiency of
S0's and an overabundance of spirals in CL0939.7+4713.  The breakdown for
spirals and irregulars in present-day clusters are 20\%, 16\% and 4\% for
Hubble types Sa, Sb and Sc+Irr (Whitmore, Gilmore and Jones 1993).
However, in CL0939.7+4713, the fractions are 12\%, 23\% and 23\%. There is
a deficiency of S0's and Sa's and an overabundance of Sb's and Sc+Irr's.
Using Dressler \etal color values for the blue galaxies ($g-r<1.21$), only
three were classed as E or S0 from the HST images, the remaining 33 are
late-type systems Sa through Irr. Of the red galaxies, 39 were E/S0 and 25
were disk systems. There were no galaxies classed later than Sc in the red
sample and the spiral sequence itself also shows a division from blue to
red with only 3 blue Sa versus 9 red Sa's, 11 blue Sb's versus 12 red Sb's
and 7 blue Sc's versus 4 red Sc's. This distribution has all the
signatures of a gas depletion scenario since the ratio of HI mass to
luminosity ($M_{HI}/L_B$) is also a decreasing function with galaxy type
such that early-type spirals have lower current star formation
rates and, therefore, redder optical colors.  However, the data from
Dressler \etal do not find large numbers of Sa's evolving into a
population of S0's but, instead, an overabundance of late-type spirals
whose B/D's are incompatible with conversion to S0's. Although there are
proposals to build large B/D galaxies from late-type galaxies with recent
star formation (see Pfenniger \etal 1994) the stellar populations of
present-day bulges do not support this hypothesis. In addition, the
luminosity of Butcher-Oemler galaxies is not much brighter than normal
cluster spirals and any fading to an S0 would produce a luminosity
function for present-day cluster S0's that is fainter than field S0's,
which is not found. All this would argue against a scenario where blue
galaxies are transformed into present-day S0's; however, a B/D study of
the HST images would further resolve this problem.

Our interpretation that the blue narrow band colors are due to disk
systems with normal star formation rates or post-starburst galaxies,
rather than an ongoing starburst event, is also confirmed by HST imaging.
Dressler (1993) reports that the brightest blue cluster galaxies are
interacting or merger systems; however, a majority are normal disk systems
in appearance. In addition, the morphology of red objects at high redshift
is homogenous (Couch \etal 1993) such that they can not be highly reddened
starbursts systems, put represent smooth, old population objects.  The
colors of the late-type galaxies are also bluer than their present-day
counterparts.  After taking the Dressler \etal $g-r$ colors, applying a
mean $K$-correction of 1.19 (variation by type was less than 0.05) and
converting to $B-V$, we obtain means colors for Sa, Sb and Sc+Irr to be
0.70, 0.60 and 0.19.  Oemler (1991) reports that cluster values for Sa, Sb
and Sc are 0.94, 0.87 and 0.69 respectively. The field values, also from
Oemler (1991), are 0.77, 0.69 and 0.52. So even at a redshift of 0.4, the
late-type galaxies have higher star formation rates then cluster spirals
today, more in-line with field spirals colors. The Sc+Irr colors are too
blue even for field spirals, but are similar to the colors of LSB disk
galaxies (McGaugh 1992). Color evolution is not confined to just the
Butcher-Oemler galaxies, but all galaxy types in the cluster.

Interestingly enough, when combining the above information from
ground-based photometry and HST imaging, none of the three evolutionary
scenarios for the blue cluster population are without significant
drawbacks or contradictions.  The fading and destruction scenarios fail to
match current observational limits for LSB galaxies in clusters or large
luminous halos.  However, if the HST results for CL0939.7+4713 are
indicative of all high redshift clusters, then neither is it plausible for
a large fraction of the blue cluster galaxies to evolve into S0's since
the progenitors are mostly small B/D late-type spirals. On the other hand,
some fading must occur since the colors of the spirals in CL0939.7+4713
are much bluer than present-day cluster spirals and, when this episode of
star formation ends, their mean surface brightnesses must decrease.
If we eliminate the late-type spirals in this manner than the ratio of
E/S0/Sa is roughly similar to present-day clusters implying that S0's had
completed their star formation by redshift of CL0939.7+4713 ($z=0.4$) in
agreement with the disk ages from Bothun and Gregg (1990). The fraction of
ellipticals is slightly higher as compared to present-day values (37\%
versus 20\%); however, bright ellipticals in rich clusters are merger
products (Schombert 1988) and, therefore, their numbers will be decreased
as dynamical friction produces numerous mergers among bright ellipticals
in the cluster core.

An alternative to the above scenarios is to assume that shortly after the
bright, star-forming period of its life, the blue cluster galaxies are
destroyed while the stellar remnants evolve and dim, a hybrid of the
destruction and fading scenarios for the evolution of the blue cluster
population. If the blue galaxies have their origin as an infalling
population of LSB disk galaxies, who are undergoing an enhanced phase of
pressure-induced star formation (Evrard 1990), then shortly
after their blue phase they will encounter the cluster core and be tidally
disrupted, spreading the fading stellar population into the cluster
potential. Since the population is undergoing a burst of star formation, it
has an enhanced surface brightness to increase their detection at
higher redshifts. In addition, the surface mass density is still as low as
their present-day field counterparts making them more susceptible to
tidal effects than normal disk systems. This hypothesis has the advantage
of preferentially destroying these low surface density blue galaxies so as
to explain why no giant LSB systems are seen in present-day clusters and
also fading the remnant stellar population that would make up the large
cluster halos.  There is some observational support for this hybrid
scenario since HST images of high redshift blue galaxies indicates that
many have a LSB appearance (Couch \etal 1993) and LSB galaxies have also
been offered up as candidates for the faint blue field population that
plagues galaxy counts studies (McGaugh 1994).  The major drawback to this
scenario that the combined effects of ram pressure induced star formation,
enhanced AGN activity from increased fuel supply, fading then tidal
destruction is a somewhat contrived scenario and too highly fine tuned to
produce large numbers of red clusters galaxies at the present epoch.

\subsection{COLOR EVOLUTION IN ELLIPTICALS}

Ellipticals are the best subjects for testing color evolution since they
are relatively free of dust, nebular emission and non-thermal sources.
They also represent the simplest test cases since all indications are that
they are composed of a single burst system, i.e. a galaxy where all the
stars formed at a single epoch of star formation near the time of 
formation and where subsequent supernovae removed the remaining gas by
galactic winds. The color history of single burst objects becomes an
exercise of composite stellar evolution. The variables in evolutionary
models of this type are the form of the stellar mass function (IMF),
metallicity distribution and history, redshift of galaxy formation and
cosmological parameters such as $H_o$ and $\Omega_o$ (see Buzzoni 1989).

An analysis of color evolution in ellipticals first requires a separation
of star-forming, AGN and other active galaxies from the ``passive"
ellipticals. Our filter system uses multiple colors to distinguish blue
from red galaxies and, therefore, a separate analysis can be done on the
red objects as individuals. At low redshifts, cluster populations are
dominated by red objects which are a mixture of ellipticals (20\%) and
S0's (40\%) (Oemler 1991). At higher redshifts, the Butcher-Oemler effect
begins to strongly influence the cluster population such that a over 80\%
of the cluster population is involved in some current (weak to strong)
star formation.  However, analysis of the stellar populations in
present-day ellipticals indicates that their epoch of star formation was
at least 4 Gyrs before $z=0.9$ and that the duration of star formation was
less than one Gyr to explain the metallicity dispersion (Rose 1985, Wyse
1985). Thus, this remaining 20\% of red galaxies at $z=0.9$ must be the
progenitors of the current epoch cluster ellipticals.

In our earlier papers we separated the ellipticals from other cluster
members with a multiple color criteria (primarily $bz-yz>0.2$,
$mz>-0.2$).  This criteria, based on the colors of present-day ellipticals
and spirals, is conservative and our analysis was relativity insensitive
to our selection process since only 20\% or less of a cluster population
was excluded up to redshifts of 0.4. However, beyond a redshift of 0.4,
the number of blue galaxies increases rapidly introducing contamination
problems from galaxies within the cluster itself. Also, by redshifts of
0.7, the mean color of an elliptical, as predicted by SED models,
approaches our blue limit for galaxy formation redshifts between 5 and
10.  The end result is that our selection of the red population will be
model dependent beyond $z=0.6$.

There are three avenues for analyzing the red population given the
expected changes in mean color from the standard models. The first is to
continue to apply our previous color selection based on present epoch
galaxies.  These produce the values shown as open symbols in Figure 6 and,
even applying this crude selection, the mean color of red objects, i.e.
ellipticals, changes by 0.2 mags bluer from a redshift of 0.5 to 0.9
whereas the changes below $z=0.5$ were very small. This is the first
evidence in our filter system that there is significant color evolution in
ellipticals.  A second method is to use the predictions of the
spectroevolutionary models to estimate a new color correction for each
redshift. In this case we have use the UV-cold models of Guiderdoni and
Rocca-Volmerangre (1987, hereafter GRV) for a redshift of formation of 5
(the best fit to the $z<0.4$ uncorrected clusters). Our original color
criteria was 0.17 blueward of the mean elliptical color at the present
epoch and we use the models to maintain this 0.17 mags difference at each
redshift. The cutoff $bz-yz$ values are listed in Table 1, column 5 and
the new mean colors using these cutoff values are found in columns 6 and 7
and plotted as solid symbols in Figure 6. The third method was to use a
constant percentage of the cluster population by color. Guided by the
morphological mixture in present-day cluster, we choose 20\% of the
reddest objects for each cluster beyond $z=0.6$.  This selection produced
a distribution of mean colors identical to the model cuts above and,
therefore, are not shown in Figure 6.

The mean colors of the red population from model cuts are shown in Figure
6 as the solid symbols (error bars are errors on the mean value for each
cluster) along with the GRV models ($H_o= 50$, $\Omega_o=0$ and a
Miller-Scalo IMF for a single 1 Gyr star burst at $z_g=5$ and 10).
Independent of the model tracks, there is a clear trend for increasing red
colors to a redshift of 0.4 with a sharp blueward change from 0.6 to 0.9.
There is little difference between the various methods of calculating the
mean cluster color out to redshifts of 0.7.  There is also fairly good
agreement with the GRV models in both $bz-yz$ and $vz-yz$ including the
prediction of a red bump at $z=0.4$ (see Paper II). This small peak is due
to a increased contribution from AGB stars 4 Gyrs ago and is not predicted
by other models which do not include late stages of stellar evolution. This
provides an example of how color observations can be used to test specific
parameters in SED models, such as stellar tracks or global metallicity.
On the other hand, a faster decrease in color is seen at high redshifts
then predicted by the models.  Interpreting the data strictly within the
context of these models indicates a redshift of elliptical formation in
clusters cores as between 4 and 5.  However, we note that from redshifts
of 0.6 to 0.9 there still exist truly red ($b-y > 0.35$) objects in all
clusters and, assuming that these are not dust shrouded starbursts, this
implies that cluster ellipticals did not all have the same epoch of
formation (see \S3.4).

The $bz-yz$ colors track the UV-cold models quite well for a formation
redshift between 4 and 5. The $vz-yz$ colors are also in good agreement
with these same models and formation redshifts. However, for individual
cluster values of $bz-yz$, the $vz-yz$ colors are slightly bluer than
model predictions by 0.05 to 0.10 mags between z=0.4 and z=0.9, although
in agreement from z=0 to 0.4. There are two possible sources for this
discrepancy between the continuum colors and the metallicity colors. One
is that there is an increasing contribution from low metallicity stars are
a function of redshift that is not accounted for in the SED models. Since
the models are only for solar metallicity stars, a distribution of
metallicities (similar to the bulge to halo distribution in our Galaxy) was
not considered. The metallicity effects of line blanketing and the mean
temperature of the giant branch as well as unusual stellar types from low
metallicity populations such as blue horizontal branch stars are more than
sufficient to cause this discrepancy and future SED codes that contain a
full chemical evolution treatment can address these observations. The
second possibility is that there is contamination from the blue galaxy 
population. The fact that the bluer $vz-yz$ colors begins at the same
redshift where the blue cluster population begins to dominate is
suggestive. In previous papers, we have pointed out that our study only
samples the bright end of the luminosity function (down to about 1/2
$L^*$).  Normally for a cluster population, the bright end is dominated by
ellipticals and S0's. However, if the blue cluster population is a
population of star-forming S0's, then some post-starburst S0's may
contribute to the mean cluster colors regardless of our selection criteria
since the distinction between blue and red galaxies is not discrete. On
the other hand, one expects a population of blue S0's to fade very quickly
since present-day cluster S0's have large B/D ratios and when undergoing a
star formation phase the disk light will dominate, but as the star
formation ceases, the large bulge component will quickly replace the disk
as the dominate source of integrated color. In either case, we can not
ignore the possibly of contamination resulting in slight variations in the
different colors with redshift.

\subsection{EPOCH OF GALAXY FORMATION}

The redshift of galaxy formation is a critical constraint on cosmological
models. For example, standard CDM with a biasing factor of $b=2.5$ 
is unable to reproduce large scale structure in a top-down hierarchy
with a galaxy formation redshift greater than 3. Early formation redshifts
($z\geq10$) have been suggested in numerous high redshift studies
(Hamilton 1985, Steidel and Hamilton 1993, Hu and Ridgeway 1994). Their
results can be summarized that up to redshifts of 3.5 one can still
identify red, old population galaxies. Given our current understanding of
the photometric evolution of a single burst population, a red galaxy at
redshifts greater than three or, 15 Gyrs ago, requires an addition two Gyrs
to evolve a dominate red giant branch. This additional evolution requires
the epoch of first star formation to be greater than $z=5$.

To explore the earliest epoch of galaxy formation the concept of the ``red
envelope'' was invented (O'Connell 1987). This idea selects out the
reddest (i.e. oldest) objects at any particular epoch then traces the red
edge in the color distribution of galaxies as a function of redshift.
Since objects in this study are selected on a multicolor plane, it is
unlikely that our red population is contaminated by highly reddened
objects but must represent objects which are red due solely to their
stellar population. To estimate the redshift of galaxy formation for
ellipticals in clusters, we define the red envelope as the 3$\sigma$ edge
of the red galaxy distribution in color.  The data for the mean colors for
red galaxies (i.e. ellipticals) in Figure 6 indicates a model dependent
formation redshift approximately five. However, the red envelope data,
shown in Figure 7 along with GRV SED models of various formation
redshifts, indicates a a formation redshift of 10 for the oldest
ellipticals in cluster cores.  At $z=1$, the typical elliptical has mean
color indicative of a F star population (E+F type rather than E+A for
post-starburst blue galaxies). Thus, we can rule out formation epochs less
than $z=4$ for ellipticals.

This estimate assumes the reddest objects are the oldest objects, but
there are several other factors which can effect the red envelope such as
systematic errors in the photometry (i.e.  bias towards detecting red
objects) or metallicity effects (i.e. galaxies with the highest mean
$[Fe/H]$ have the reddest integrated colors).  Systematic errors in our
photometry seem unlikely and work in the opposite direction since sky
brightness increased in our reddest filters enhancing the detection of
blue galaxies. Metallicity effects are also minor since the
color-magnitude relation indicates very small changes in color (less than
0.002 mags in $bz-yz$, see Schombert \etal 1993) over the range of
luminosity sampled herein ($L > 1/2L^*$). However, we note that the
$vz-yz$ colors are slightly bluer than the $bz-yz$ colors compared to
model tracks and indicate a redshift of formation of 8. Since a
metallicity distribution of stars within the galaxy was not specifically
included in the models (see Arimoto and Yoshii 1987), then the
interpretation of red envelope is more accurate using $bz-yz$ colors.  In
addition, since the red envelope indicates a redshift of formation of 10,
yet the mean colors produce a value of 5, we interpret this to imply that
cluster ellipticals are not coeval.

In the same fashion as the red envelope, we can define a blue envelope to
estimate the formation epoch of the blue cluster population. Rather then
defining a lower 3$\sigma$ envelope, we simply calculate the mean color of
the blue population as shown in Figure 7. The track from these colors
indicate a formation redshift between two and three. If the blue
population is composed of LSB galaxies, then this would correct
identify this first epoch of star formation, even though the galaxy itself
may have formed as a quiescent gas cloud at higher redshifts. This scenario is
consistent with the spatial distribution of LSB galaxies (Mo \etal 1994)
and quiescent gas clouds (Lacy \etal 1993). If the blue population is
composed of large B/D proto-S0's, then this track would not indicate the
formation redshift, since the spheroidal components have similar ages to
ellipticals, but rather the last epoch of disk star formation at $z=0.4$, a
redshift consistent with S0 disk age estimates (Bothun and Gregg 1990).

\section{SUMMARY}

Although our understanding of color evolution and the formation of
galaxies has focused on searches for distant protogalaxies, there is
still a great deal that can be learned from photometry of lower redshift
clusters from $z=0$ to $z=1$. This study has taken several small steps 
with information extracted from simple photometry using narrow band
filters that sample specific portions of the spectrum relevant to star
formation, metallicity and other evolutionary factors. We adjust the
center wavelength of the filters to match the rest frame of the cluster,
thus avoiding any model dependent K-corrections or strong emission lines.
We summarize our primary results as the following:

\begin{itemize}\begin{enumerate}
\item We find a continuation of the Butcher-Oemler effect, the
fraction of blue to red galaxies in a cluster, from values of $f_B=0.20$
at redshifts of 0.4 to $f_B=0.80$ at redshifts of 0.9. This trend is much
stronger than found by early studies and suggests that all cluster
galaxies with disks (S0's and spirals) are involved in star formation 9 to
10 Gyr's ago.

\item The colors of the Butcher-Oemler galaxies, the blue cluster
population, is similar to that of an ongoing and post-starburst
population.  Redder $uz-vz$ colors (a strong 4000\AA\ break) indicates a
young stellar population similar to the E+A galaxies found
spectroscopically by Dressler and Gunn (1985). The formation epoch of this
population is between a redshift of 2 and 3 based on a blue envelope
analysis.

\item Our color information, when combined with morphological HST
studies in the literature, leads us to conclude that the evolution of the
blue cluster population is such that they can not be converted to red
galaxy types (i.e. ellipticals and S0's) on the short timescales from
$z=0.5$ to the present. We propose that the Butcher-Oemler effect is due
to a population of low surface density galaxies which are blue due to
pressure induced star formation that occurs upon infall to the cluster
environment. This population is later preferentially destroyed by the
cluster tidal field over red galaxy types and the stellar remnants fade
below detectable limits.

\item The mean colors of ellipticals displays a trend of redder
colors until a redshift of 0.4 then a blueward drop out to the edge of the
sample at $z=0.9$. We find good agreement with the UV-cold SED models of
Guiderdoni and Rocca-Volmerangre (1987) including a red bump at $z=0.4$
due to an AGB star contribution and a smooth transition to blue optical
colors by a redshift of 0.9. Our metallicity color, $vz-yz$, is slightly
bluer than model predictions beyond a redshift of 0.6, probably due to a
metal-poor component to the underlying stellar population and indicates
the need for a full chemical evolutionary treatment in the models 
(see Arimoto and Yoshii 1987).

\item Assuming the cosmological constants needed to match
globular cluster ages, we find the mean epoch of galaxy formation to be
$z=5$. However, tracing the red envelope in the cluster data finds the
oldest (i.e. reddest) galaxies to have formation redshifts of
approximately 10. This is a major challenge to structure formation
theories.
\end{enumerate}\end{itemize}

\acknowledgements
We wish to thank the generous support of KPNO, Lowell Obs. and the Univ.
of Michigan during the last three years of collecting the high redshift
data. Special thanks to Barry Madore for correctly pointing out the
meaning of our red $uz-vz$ colors. The research described herein was
carried out by the Jet Propulsion Laboratory, California Institute of
Technology, under a contract with the National Aeronautics and Space
Administration. Additional financial support from the Austrian Fonds zur
F\"orderung der Wissenschaftlichen Forschung is gratefully acknowledged.

\figcaption[] {The modified Str\"omgren filter system used
herein is shown along with the SED models for an old, metal-rich galaxy
(17 Gyr) and a young post-starburst SED (1.5 Gyrs, from Guiderdoni and
Rocca-Volmerangre 1987). Note that although the near-UV flux increases by
any measure of $uz-yz$, $vz-yz$ or $bz-yz$, the 4000\AA\ break colors,
$uz-vz$, become redder with a younger population.\label{fig1}}

\figcaption[] {A greyscale image of our most distant cluster,
CL1622.6+2352, taken with the KPNO PFCCD system for 3000 secs in our $yz$
filter. The field of view is 3.2 arcmin with north at the top and east to
the left. The cluster is the small concentration of faint objects in the
center of the frame. (This figure is not included in this preprint version) \label{fig2}}

\figcaption[] {Histograms of the continuum color, $bz-yz$ for
the 17 clusters in our samples. Data for clusters with redshifts less than
0.6 are from Papers I through III. Note how the median color shifts by 0.7
mags from $z=0.2$ to 0.9.\label{fig3}}

\figcaption[] {The Butcher-Oemler effect as seen in rest frame
Str\"omgren colors. Solid diamonds are the new data from this paper, the
crosses are from the original Butcher and Oemler (1985) data.  We confirm
previous studies in that strong evolution is present in the data. The
dashed line is the original interpretation of the Butcher-Oemler effect.
The dot-dash line is our new estimation based on the higher redshift
data.  The fraction of blue galaxies ($f_B$) rises sharply from a
present-day value of 0.04 to approximately 0.80 at a redshift of 0.9 in
roughly a linear fashion. Since the mixture of present-day galaxy types is
20\% E's, 40\% S0's and 40\% spirals in rich clusters, this new data
indicates that all disk systems are consumed in significant star formation
8 Gyrs ago.\label{fig4}}

\figcaption[] {Histograms of near-UV color, $uz-yz$, and the
4000\AA\ break color, $uz-vz$. The top panels displays the summed data for
all cluster galaxies with a redshift less than 0.6. The second panels
display the summed data for all cluster galaxies with redshifts greater
than 0.6.  The bottom sets of panels display the distribution of colors
for red and blue galaxies (selected by $bz-yz$ colors) from only the high
redshift clusters. The near-UV color changes little between the high and
low redshift samples, although the blue and red galaxies are clearly
distinguished by UV flux, assumingly from increased massive star
contribution in the blue galaxies. The blue and red galaxies are identical
in $uz-vz$, i.e. the blue galaxies are too red for their continuum
colors.  This signals an enhanced 4000\AA\ break component from a young
stellar population.\label{fig5}}

\figcaption[] {Color as a function of redshift for the red
population in each cluster (mean values, error bars are errors on the
mean). These objects are the progenitors of the present-day cluster
ellipticals.  Open symbols are raw colors using a constant $bz-yz<0.2$
cutoff. Solid symbols are mean colors from model determined cutoffs listed
in Table 1.  Similar trends are seen in both our metallicity color ($v-y$)
and continuum color ($b-y$). Evolution in both filters is roughly redward
till $z=0.5$ then a sharp blueward trend to $z=1$. Model tracks are from
Guiderdoni and Rocca-Volmerange (1987) for the cosmology and star
formation parameters listed. Good agreement exists between the model and
observations, although the metallicity color is slightly bluer than model
predictions matched to $bz-yz$ color indicating a low metallicity
population unaccounted for by the models.\label{fig6}}

\figcaption[] {Red and Blue envelopes to determine the
earliest epoch of galaxy formation and the star formation epoch of the blue
cluster galaxies.  The reddest galaxies are selected from 3$\sigma$ from
the cluster mean colors to form the red envelope. Model fits suggest the
earliest redshift of elliptical formation is at least 10. The median
colors of the blue cluster population are used to define the blue
envelope. The blue envelope indicates the epoch of star formation for the
Butcher-Oemler effect is between a redshift of 2 and 3.\label{fig7}}

\clearpage
\begin{figure}
\epsscale{1.3}
\plotfiddle{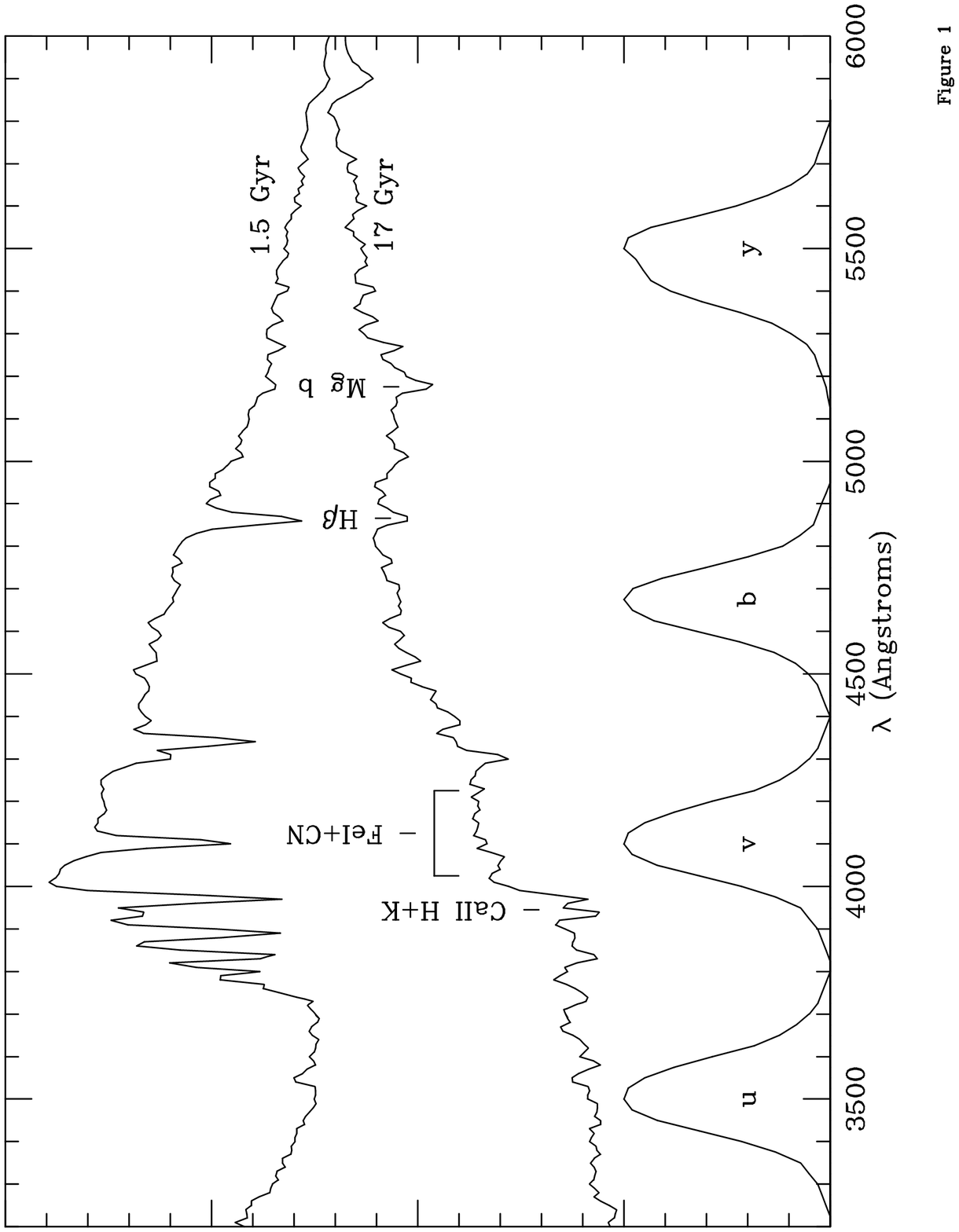}{11.5truein}{0}{90}{90}{-100}{210}
\end{figure}
\clearpage
\begin{figure}
\epsscale{1.3}
\plotfiddle{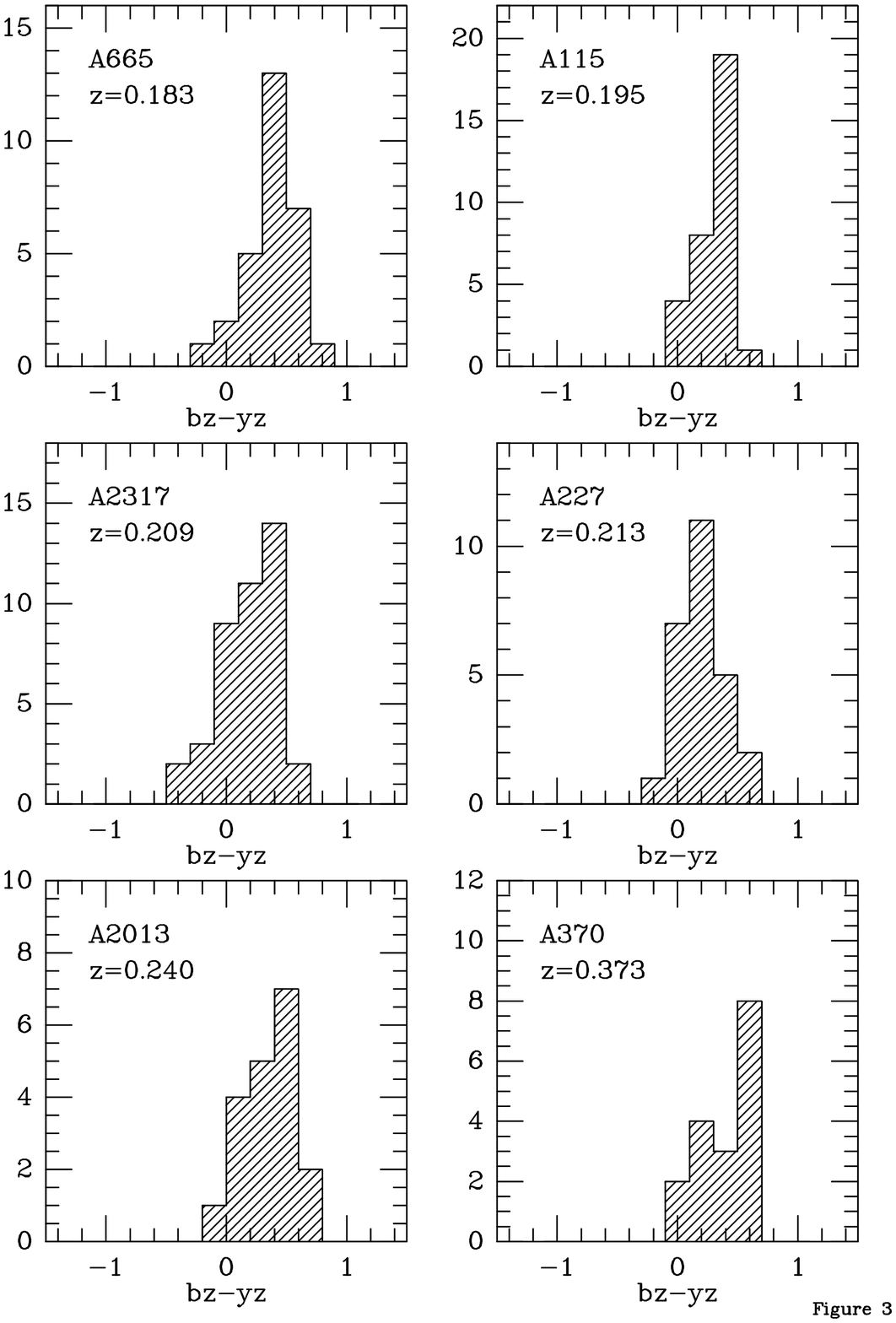}{11.5truein}{0}{90}{90}{-150}{210}
\end{figure}
\clearpage
\begin{figure}
\epsscale{1.3}
\plotfiddle{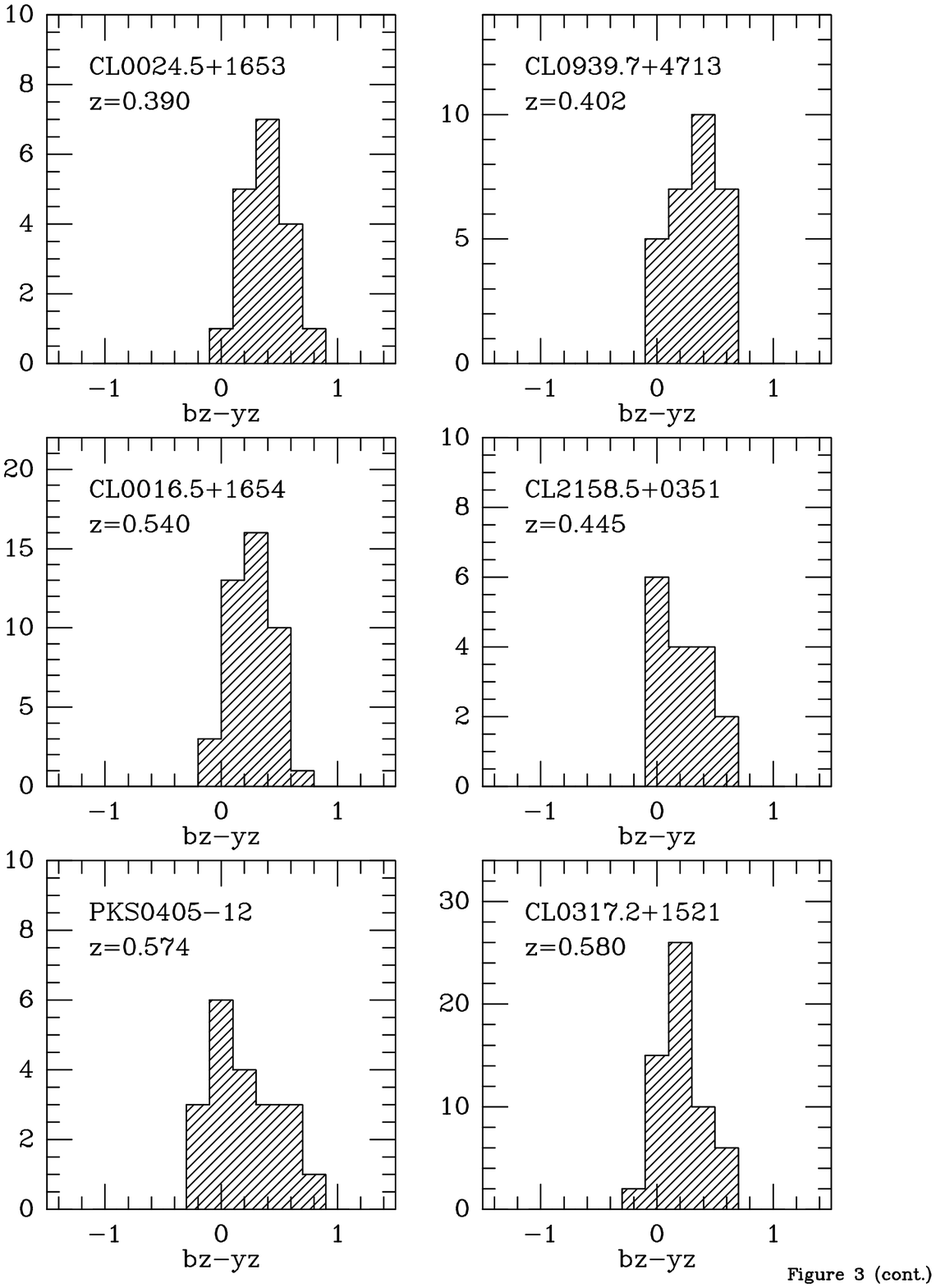}{11.5truein}{0}{90}{90}{-150}{210}
\end{figure}
\clearpage
\begin{figure}
\epsscale{1.3}
\plotfiddle{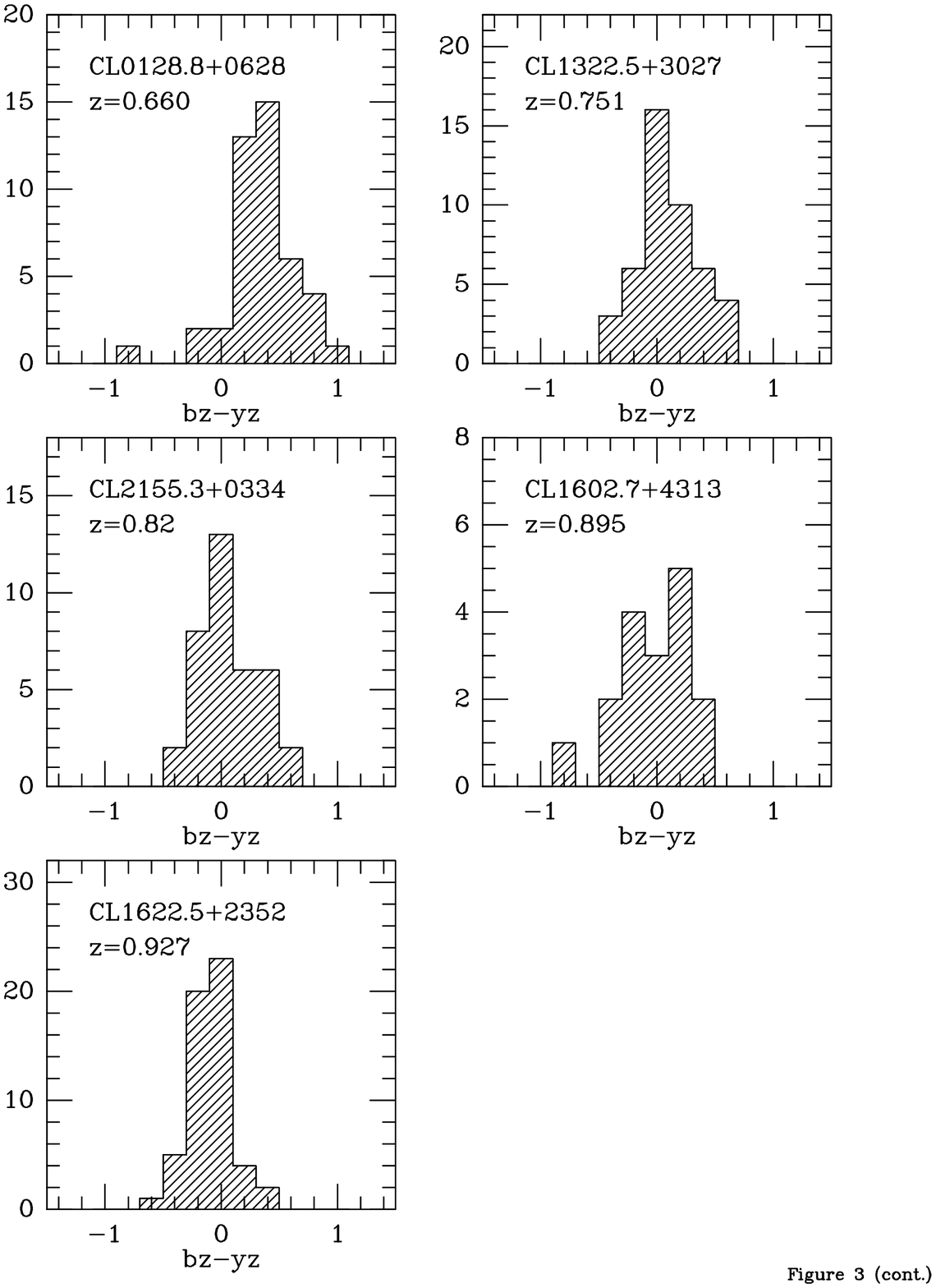}{11.5truein}{0}{90}{90}{-150}{210}
\end{figure}
\clearpage
\begin{figure}
\epsscale{1.3}
\plotfiddle{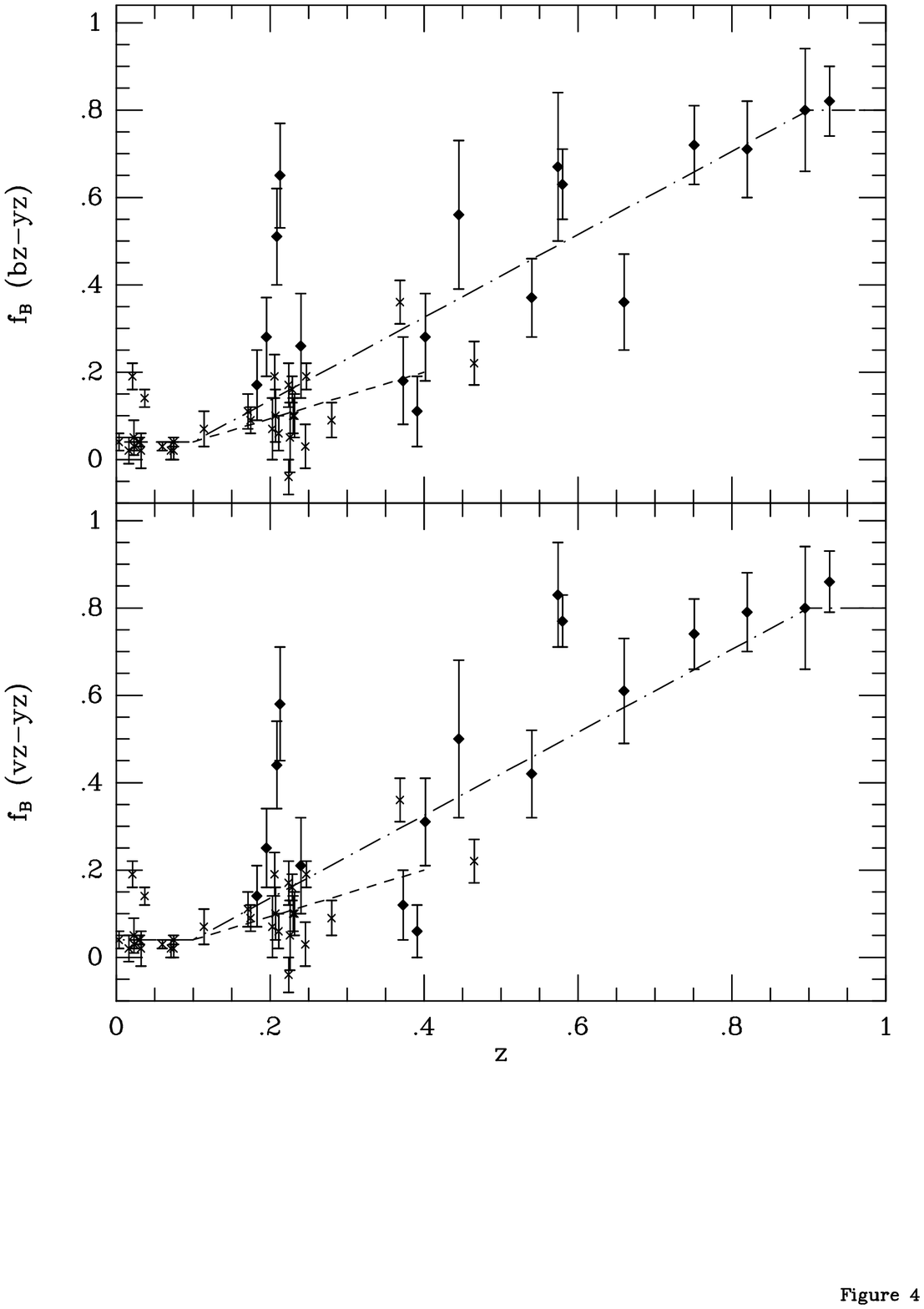}{11.5truein}{0}{90}{90}{-150}{210}
\end{figure}
\clearpage
\begin{figure}
\epsscale{1.3}
\plotfiddle{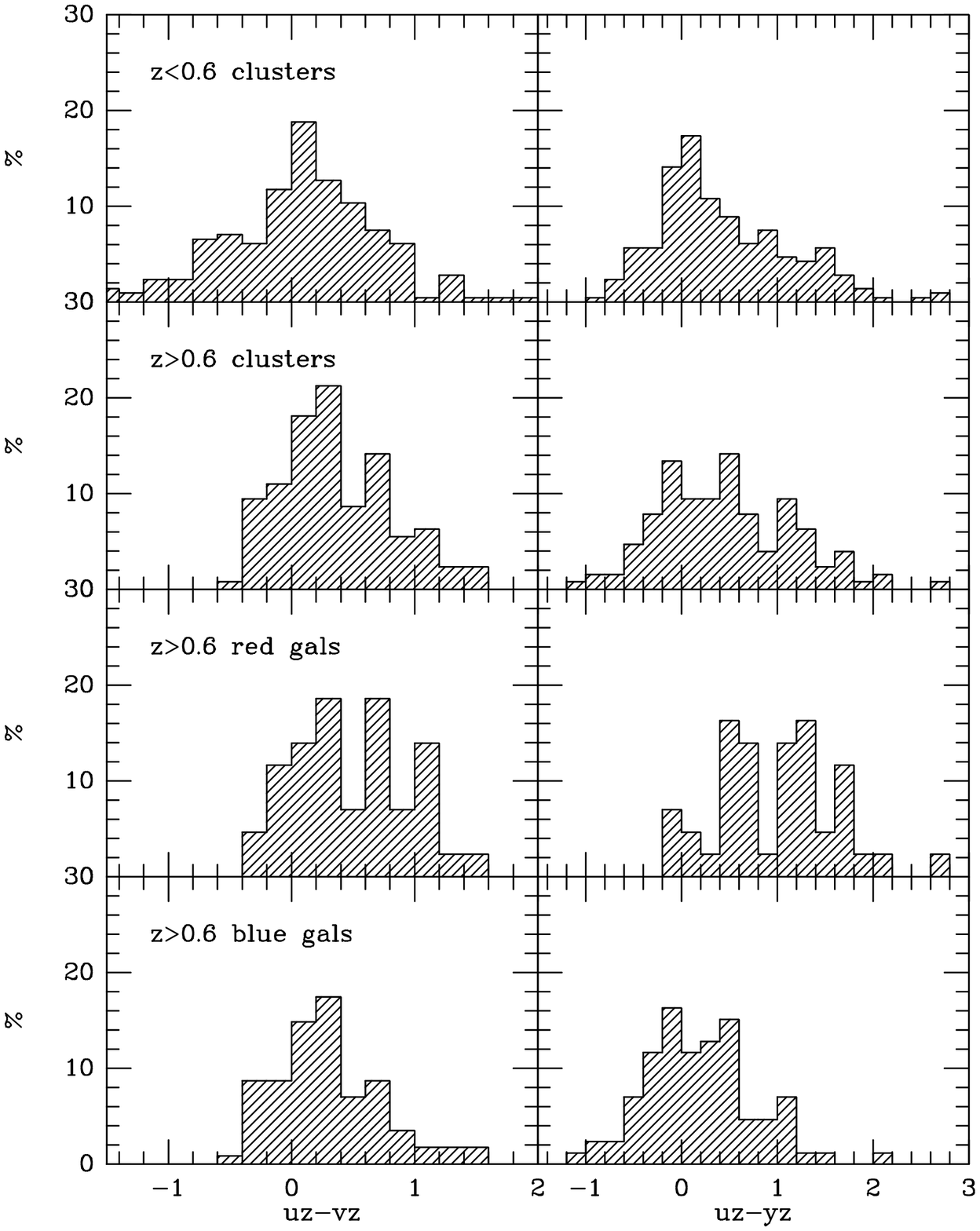}{11.5truein}{0}{90}{90}{-150}{210}
\end{figure}
\clearpage
\begin{figure}
\epsscale{1.3}
\plotfiddle{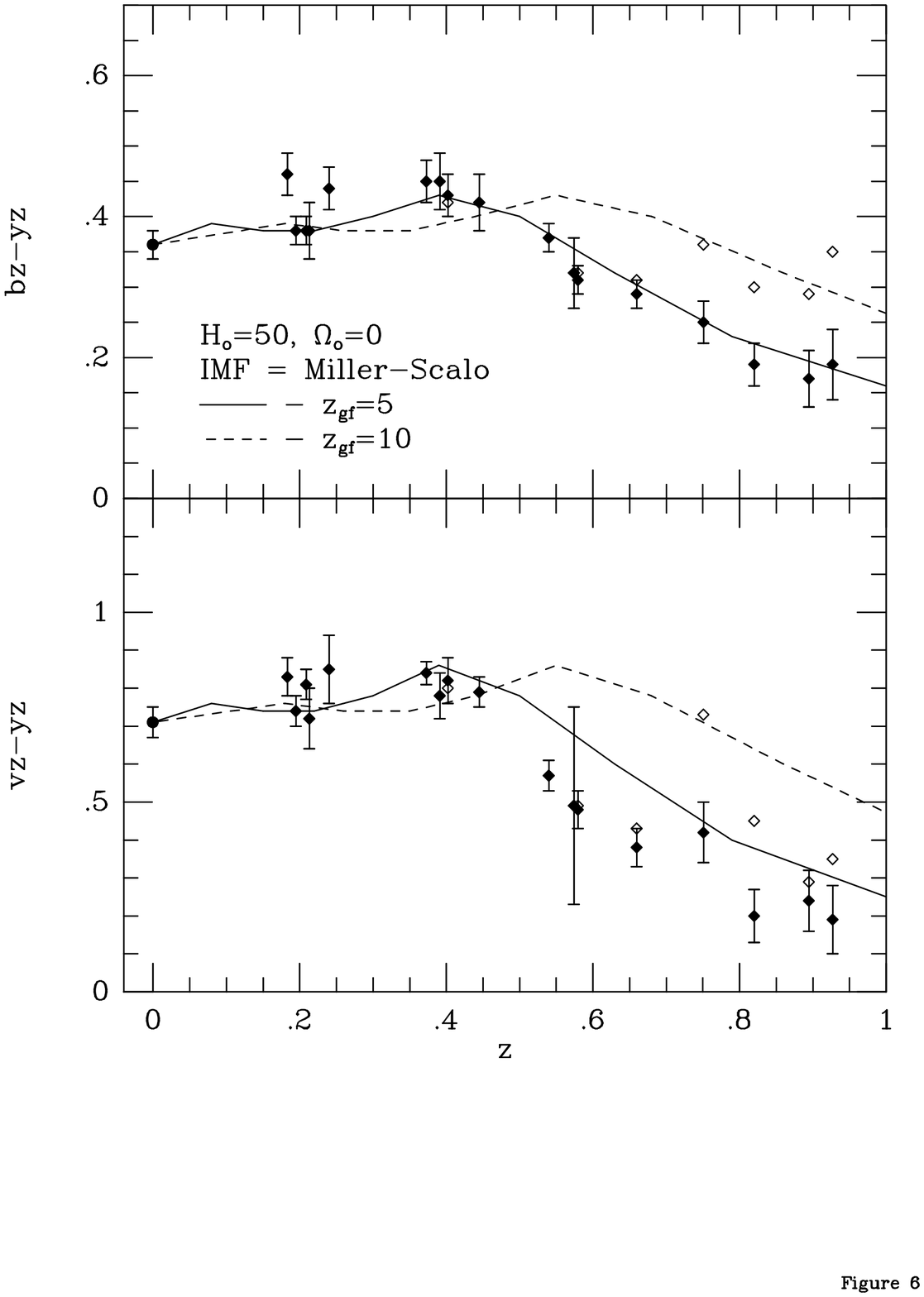}{11.5truein}{0}{90}{90}{-150}{210}
\end{figure}
\clearpage
\begin{figure}
\epsscale{1.3}
\plotfiddle{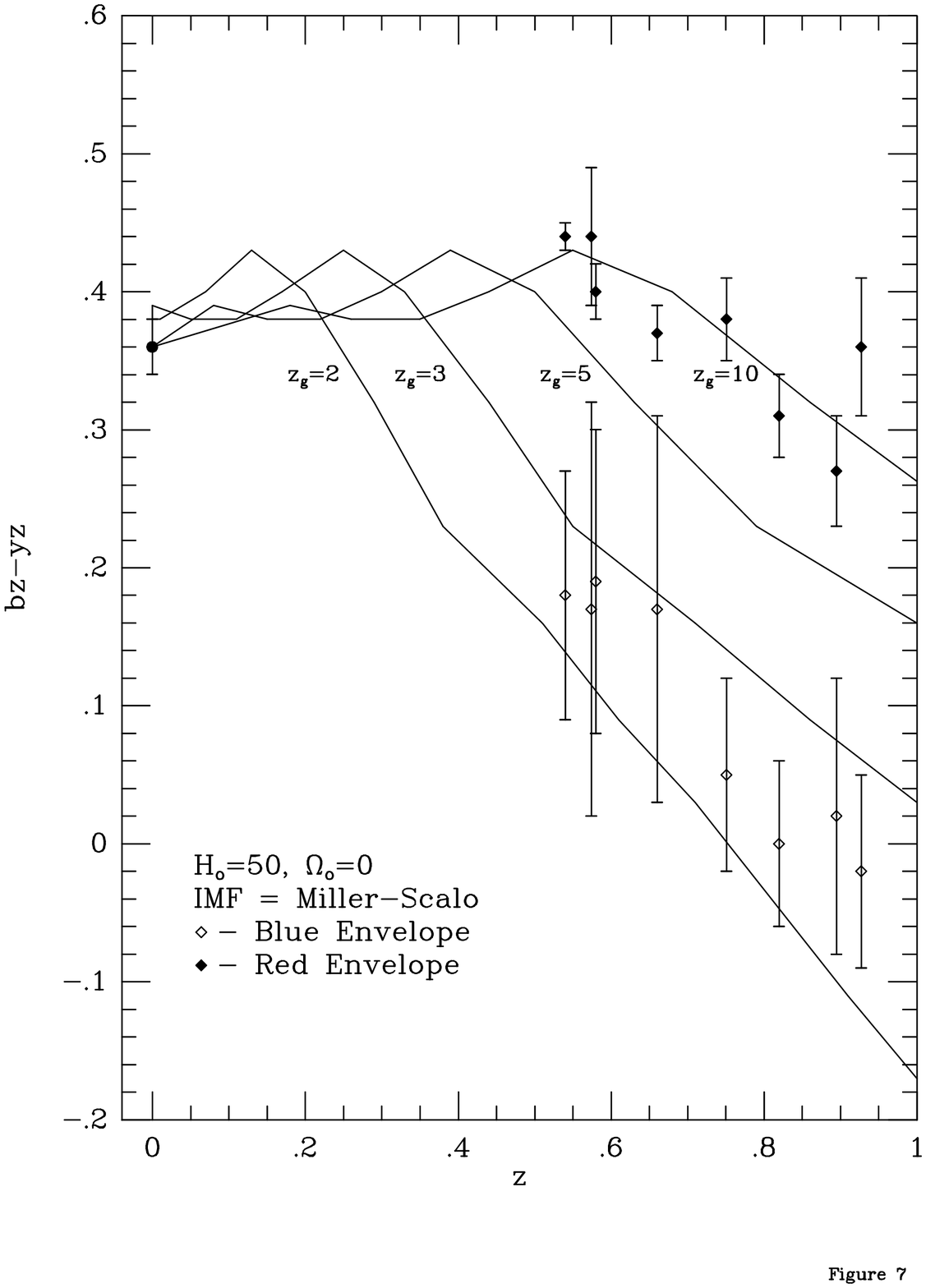}{11.5truein}{0}{90}{90}{-150}{210}
\end{figure}

\begin{references}
\reference{} Allington-Smith, J., Ellis, R., Zirbel, E. and Oemler, A. 1993,
\apj, 404, 521
\reference{} Arimoto, N. and Yoshii, Y. 1987, \aap, 173, 23
\reference{} Binggeli, B., Sandage, A. and Tammann, G. 1985, \aj, 90, 1681
\reference{} Bothun, G., Schombert, J., Impey, C., Sprayberry, D. and McGaugh, S. 1993, \aj, 106, 530
\reference{} Bothun, G. and Gregg, M. 1990, \apj, 350, 73
\reference{} Butcher, H. and Oemler, A. 1984, \apj, 285, 426
\reference{} Buzzoni, A. 1989, \apjs, 71, 817
\reference{} Couch, W., Ellis, R., Sharples, R. and Smail, I. 1993, {\it
Observational Cosmology}, eds: G. Chincarini, A. Iovino, T. Maccacaro and
D. Maccagni, A.S.P. Conference Series, Vol. 51, p. 240
\reference{} Ellis, R. 1988, in {\it Cooling Flows in Clusters and Galaxies}, ed. A.
Fabian, Kluwer Academic Publishers, p. 305
\reference{} Evrard, A. 1990, \apj, 363, 349
\reference{} Dressler, A., Oemler, A., Butcher, H. and Gunn, J. 1994, \apj,
430, 107
\reference{} Dressler, A. 1993, in {\it Observational Cosmology}, eds: G. 
Chincarini, A. Iovino, T. Maccacaro and
D. Maccagni, A.S.P. Conference Series, Vol. 51, p. 225
\reference{} Dressler, A., Gunn, J. and Schneider, D. 1985, \apj, 294, 70
\reference{} Dressler, A. and Gunn, J. 1983, \apj, 270, 7
\reference{} Dressler, A. 1980, \apj, 236, 351
\reference{} Eisenhardt, P. and Chokshi, A. 1990, \apj, 351, L9
\reference{} Eisenhardt, P. and Lebofsky, M. 1987, \apj, 316, 70
\reference{} Evrard, A. 1990, \apj, 363, 349
\reference{} Fiala, N., Rakos, K. and Stockton, A. 1986, \pasp, 98, {70 (Paper I)}
\reference{} Guiderdoni, B. and Rocca-Volmerangre, B. 1987, \aap, 186, {1
(GRV)}
\reference{} Gunn, J., Hoessel, J. and Oke, J. 1986, \apj, 306, 30
\reference{} Gunn, J. and Gott, J. 1972, \apj, 176, 1
\reference{} Hamilton, D. 1985, \apj, 297, 371
\reference{} Hu, E. and Ridgeway, S. 1994, \aj, 107, 1303
\reference{} Impey, C., Bothun, G. and Malin, D. 1988, \apj, 330, 634
\reference{} Knezek, P. and Schombert, J. 1994, private communication
\reference{} Lacy, C., Guiderdoni, B., Rocca-Volmerangre, B. and Silk, J. 1993, \apj,
402, 15
\reference{} Larson, R., Tinsley, B. and Caldwell, C. 1980, \apj, 237, 692
\reference{} Mamon, G. 1986, \apj, 307, 426
\reference{} Massey, P., Strobel, K., Barnes, J. and Anderson, E. 1988, \apj, 
328, 315
\reference{} Malumuth, E. and Richstone, D. 1984, \apj, 276, 413
\reference{} Matsushima, S. 1969, \apj, 158, 1137
\reference{} McCarthy, P., Persson, S. and West, S. 1992, \apj, 386, 52
\reference{} McGaugh, S. 1992, {\it Ph.D. thesis}, Univ. of Michigan
\reference{} McGaugh, S. 1994, \apj, 426, 135
\reference{} Merritt, D. 1985, \apj, 289, 18
\reference{} Mo, H., McGaugh, S. and Bothun, G. 1994, \mnras, 267, 129
\reference{} O'Connell, R. 1987, in {\it Towards Understanding Galaxies at Large
Redshift}, eds. R. Kron and A. Renzini (Dordrecht: Reidel), p.46
\reference{} Oemler, A. 1991, in {\it Clusters and Superclusters of Galaxies}, 
ed. A. Fabian, (Dordrecht: Kluwer), p. 666
\reference{} Oemler, A. 1976, \apj, 209, 693
\reference{} Pfenniger, D., Combes, F. and Martinet, L. 1994, \aap,
285, 79
\reference{} Rakos, J., Fiala, N. and Schombert, J. 1988, \apj, 328, {463 (Paper
II)}
\reference{} Rakos, K., Schombert, J. and Kreidl, T. 1991, \apj, 377, {382
(Paper III)}
\reference{} Rose, J. 1985, \aj, 90, 1927
\reference{} Schneider, D., Gunn, J. and Hoessel, J. 1983, \apj, 264, 337
\reference{} Schombert, J., Hanlan, P., Barsony, M. and Rakos, K. 1993, \aj,
106, 923
\reference{} Schombert, J., Bothun, G., Schneider, S. and McGaugh, S. 1992,
\aj, 103, 1107
\reference{} Schombert, J. 1988, \apj, 328, 475
\reference{} Spitzer, L. and Baade, W. 1951, \apj, 113, 413
\reference{} Steidel, C. and Hamilton, D. 1993, \aj, 105, 2017
\reference{} Whitmore, B., Gilmore, D. and Jones, C. 1993, \apj, 407, 489
\reference{} Wyse, R. 1985, \apj, 299, 593
\end{references}
\end{document}